\documentclass{article}
\usepackage{spconf,amsmath,graphicx}
\usepackage{spconf,amsmath,graphicx}
\usepackage{float}
\usepackage{verbatim}
\usepackage{tabularx}
\usepackage{diagbox}
\usepackage{multirow}
\usepackage{colortbl}
\usepackage{geometry}
\usepackage{soul}
\usepackage{xcolor}
\title{\vspace{0.3cm}Transcribing Lyrics from Commercial Song Audio: the First Step towards Singing Content Processing}
\name{Che-Ping Tsai$^*$\thanks{$^*$ indicates equal contribution.}, Yi-Lin Tuan$^*$\footnotemark[1], Lin-shan Lee}

\address{National Taiwan University\\
    Department of Electrical Engineering\\
    \texttt{r06922039@ntu.edu.tw,b02901048@ntu.edu.tw, lslee@gate.sinica.edu.tw}}
\begin{document}
\ninept
\maketitle
\begin{abstract}
Spoken content processing (such as retrieval and browsing) is maturing, but the singing content  is still almost completely left out. Songs are human voice carrying plenty of semantic information just as speech, and may be considered as a special type of speech with highly flexible prosody. The various problems in song audio, for example the significantly changing phone duration over highly flexible pitch contours, make the recognition of lyrics from song audio much more difficult. This paper reports an initial attempt towards this goal. We collected music-removed version of English songs directly from commercial singing content. The best results were obtained by TDNN-LSTM with data augmentation with 3-fold speed perturbation plus some special approaches. The WER achieved (73.90\%) was significantly lower than the baseline (96.21\%), but still relatively high. 
\end{abstract}
\section{Introduction}
\label{sec:intro}
The exploding multimedia content over the Internet, has created a new world of spoken content processing, for example the retrieval\cite{lee2015spoken, chelba2008retrieval, larson2012spoken, mandal2014recent, lee2014improved}, browsing\cite{lee2005spoken}, summarization\cite{lee2015spoken, lee2005spoken, shiang2013supervised, lee2013unsupervised}, and comprehension\cite{tseng2016towards, fang2016hierarchical,lee2014spoken, shen2015structuring} of spoken content. 
On the other hand, we may realize there still exists a huge part of multimedia content not yet taken care of, i.e., the singing content or those with audio including songs. Songs are human voice carrying plenty of semantic information just as speech. It will be highly desired if the huge quantities of singing content can be similarly retrieved, browsed, summarized or comprehended by machine based on the lyrics just as speech. For example, it is highly desired if song retrieval can be achieved based on the lyrics in addition.

Singing voice can be considered as a special type of speech with highly flexible and artistically designed prosody: the rhythm as artistically designed duration, pause and energy patterns, the melody as artistically designed pitch contours with much wider range, the lyrics as artistically authored sentences to be uttered by the singer. So transcribing lyrics from song audio is an extended version of automatic speech recognition (ASR) taking into account these differences.

On the other hand, singing voice and speech differ widely in both acoustic and linguistic characteristics. Singing signals are often accompanied with some extra music and harmony, which are noisy for recognition. The highly flexible pitch contours with much wider range\cite{sasou2005auto, kawailyric}, the significantly changing phone durations in songs, including the prolonged vowels\cite{sasou2006singing, kawai2016speech} over smoothly varying pitch contours, create much more problems not existing in speech. The falsetto in singing voice may be an extra type of human voice not present in normal speech. Regarding linguistic characteristics\cite{hosoya2005lyrics, mesaros2010recognition}, word repetition and meaningless words (e.g.oh) frequently appear in the artistically authored lyrics in singing voice.

Applying ASR technologies to singing voice has been studied for long. However, not too much work has been reported, probably because the recognition accuracy remained to be relatively low compared to the experiences for speech. But such low accuracy is actually natural considering the various difficulties caused by the significant differences between singing voice and speech. An extra major problem is probably the lack of singing voice database, which pushed the researchers to collect their own closed datasets\cite{sasou2005auto, kawai2016speech, mesaros2010recognition}, which made it difficult to compare results from different works.

Having the language model learned from a data set of lyrics is definitely helpful\cite{kawai2016speech, mesaros2010recognition}. Hosoya et al.\cite{hosoya2005lyrics} achieved this with finite state automaton. Sasou et al.\cite{sasou2005auto} actually prepared  a language model for each song. In order to cope with the acoustic characteristics of singing voice, Sasou et al.\cite{sasou2005auto, sasou2006singing} proposed AR-HMM to take care of the high-pitched sounds and prolonged vowels, while recently Kawai et al.\cite{kawai2016speech} handled the prolonged vowels by extending the vowel parts in the lexicon, both achieving good improvement. Adaptation from models trained with speech was attractive, and various approaches were compared by Mesaros el al.\cite{mesaros2009adaptation}.

In this paper, we wish our work can be compatible to more available singing content, therefore in the initial effort we collected about five hours of music-removed version of English songs directly from commercial singing content on YouTube. The descriptive term {\it "music-removed"} implies the background music have been removed somehow. Because many very impressive works were based on Japanese songs\cite{sasou2005auto, kawailyric, sasou2006singing, kawai2016speech, hosoya2005lyrics}, the comparison is difficult. We analyzed various approaches with HMM, deep learning with data augmentation, and acoustic adaptation on fragment, song, singer, and genre levels, primarily based on fMLLR\cite{gales1998maximum}. We also trained the language model with a corpus of lyrics, and modify the pronunciation lexicon and increase the transition probability of HMM for prolonged vowels. Initial results are reported.

\section{DATABASE}
\label{sec:data}
\subsection{Acoustic Corpus}
To make our work easier and compatible to more available singing content, we collected 130 music-removed (or vocal-only) English songs  from www.youtube.com so as to consider only the vocal line.The music-removing processes are conducted by the video owners, containing the original vocal recordings by the singers and vocal elements for remix purpose. \footnote{Samples of our collected data: https://youtu.be/QA6x9MLgsc8}

After initial test by speech recognition system trained with LibriSpeech\cite{panayotov2015librispeech}, we dropped 20 songs, with WERs exceeding 95\%. The remaining 110 pieces of music-removed version of commercial English popular songs were produced by 15 male singers, 28 female singers and 19 groups. The term {\it group} means by more than one person. No any further preprocessing was performed on the data, so the data preserves many characteristics of the vocal extracted from commercial polyphonic music, such as harmony, scat, and silent parts. Some pieces also contain overlapping verses and residual background music, and some frequency components may be truncated. Below this database is called {\bf vocal data} here.

These songs were manually segmented into fragments with duration ranging from 10 to 35 sec primarily at the end of the verses. Then we randomly divided the vocal data by the singer and split it into training and testing sets. We got a total of 640 fragments in the training set and 97 fragments in the testing set. The singers in the two sets do not overlap. The details of the vocal data are listed in Table.\ref{tab: AcousticDatabase}.

Because music genre may affect the singing style and the audio, for example, hiphop has some rap parts, and rock has some shouting vocal, we obtained five frequently observed genre labels of the vocal data from wikipedia\cite{wiki} : pop, electronic, rock, hiphop, and R\&B/soul. The details are also listed in Table.\ref{tab: AcousticDatabase}. Note that a song may belong to multiple genres.

To train initial models for speech for adaptation to singing voice, we used 100 hrs of English clean speech data of LibriSpeech.

\definecolor{Gray}{gray}{0.85}
\begin{table}[]
\centering
\begin{tabular}{|c|c|c|c|c|}
\hline
 & \cellcolor{Gray}\# songs 
 & \multicolumn{1}{c||}{\cellcolor{Gray}\# singers} 
 & pop & electronic\\
\hline
Training set 
& \cellcolor{Gray}95 
& \multicolumn{1}{c||}{\cellcolor{Gray}49} 
& 202.2  & 85.8 \\
\hline
Testing set 
&\cellcolor{Gray}15 
& \multicolumn{1}{c||}{\cellcolor{Gray}13} 
&20.3  & 22.0\\
\hline
\hline
 & rock & hiphop &\multicolumn{1}{c||}{R\&B/soul} & \cellcolor{Gray}total  \\ \hline
Training set & 51.1 &  30.0  
& \multicolumn{1}{c||}{87.5} 
& \cellcolor{Gray}271\\ 
\hline
Testing set  & 17.7 &  8.4    
& \multicolumn{1}{c||}{9.1}   
& \cellcolor{Gray}42.8 \\ 
\hline
\end{tabular}
\caption{Information of training and testing sets in vocal data. The lengths are all measured in minutes.}
\label{tab: AcousticDatabase}
\end{table}

\subsection{Linguistic Corpus}

 In addition to the data set from LibriSpeech (803M words, 40M sentences), we collected 574k pieces of lyrics text (totally 129.8M words) from {\it lyrics.wikia.com}, a lyric website, and the lyrics were normalized by removing punctuation marks and unnecessary words (like  ’[CHORUS]’). Also, those lyrics for songs within our vocal data were removed from the data set.
 
\label{subsec:approach-flow}

\begin{figure}[htb]
\begin{minipage}[b]{1.0\linewidth}
  \centering
  \centerline{\includegraphics[angle=-90, width=8.5cm]{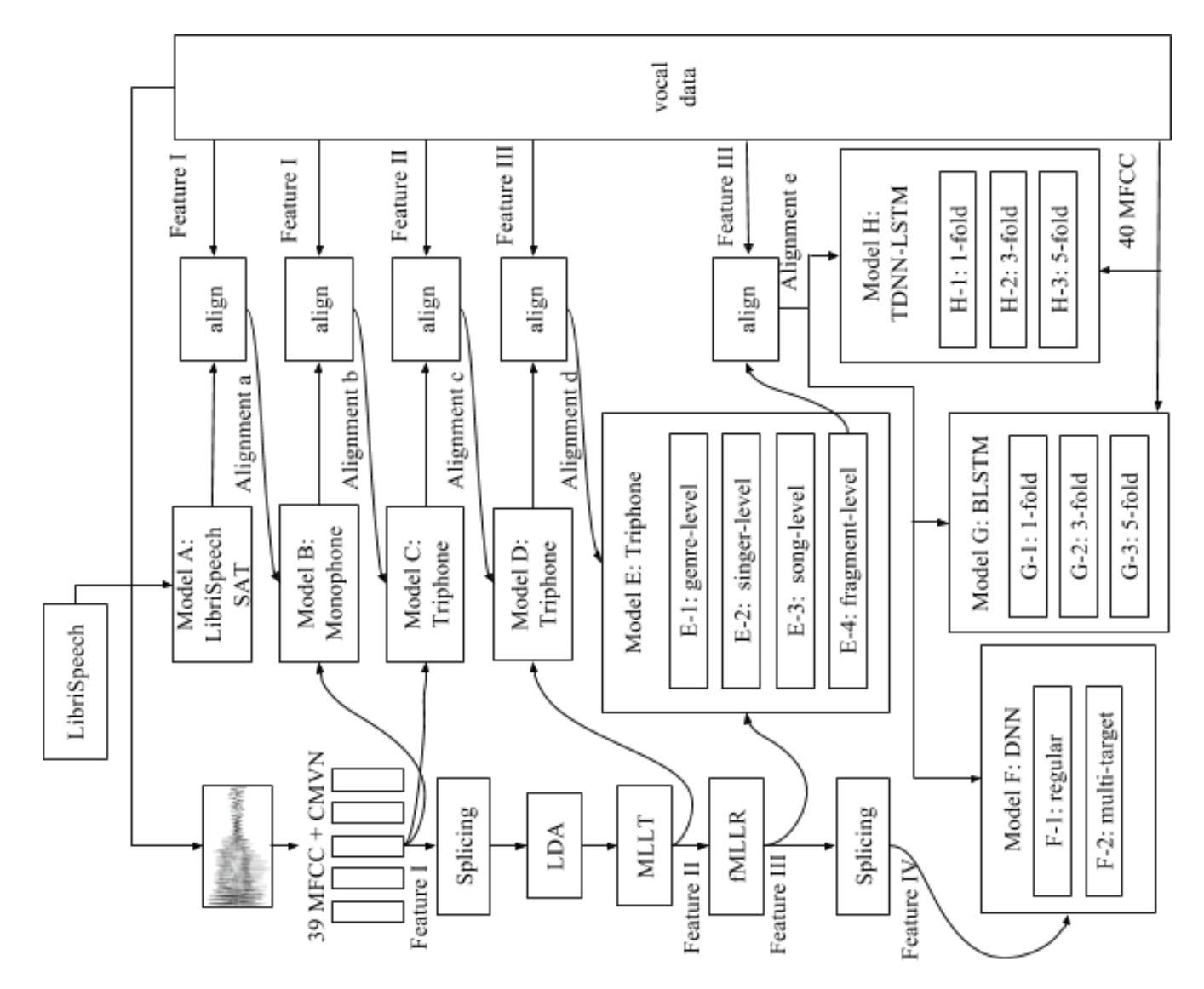}}
\end{minipage}
\caption{The overall structure for training the acoustic models.}
\label{fig:approach flow}
\end{figure}

\section{Recognition Approaches and System Structure}
\label{sec:approach}

Fig.\ref{fig:approach flow} shows the overall structure based on Kaldi\cite{povey2011kaldi} for training the acoustic models used in this work. The right-most block is the vocal data, and the series of blocks on the left are the feature extraction processes over the vocal data. Features \uppercase\expandafter{\romannumeral 1}, \uppercase\expandafter{\romannumeral 2}, \uppercase\expandafter{\romannumeral 3}, \uppercase\expandafter{\romannumeral 4} represent four different versions of features used here. For example, Feature \uppercase\expandafter{\romannumeral 4} was derived from splicing Feature \uppercase\expandafter{\romannumeral 3} with 4 left-context and 4 right-context frames, and Feature         \uppercase\expandafter{\romannumeral 3} was obtained by performing fMLLR transformation over Feature \uppercase\expandafter{\romannumeral 2}, while Feature \uppercase\expandafter{\romannumeral 1} has been mean and variance normalized, etc.

The series of second right boxes are forced alignment processes performed over the various versions of features of the vocal data. The results are denoted as Alignment a, b, c, d, e. For example, Alignment a is the forced alignment results obtained by aligning Feature \uppercase\expandafter{\romannumeral 1} of the vocal data with the LibriSpeech SAT triphone model (denoted as Model A at the top middle).

The series of blocks in the middle of Fig.\ref{fig:approach flow} are the different versions of trained acoustic models. For example, model B is a monophone model trained with Feature \uppercase\expandafter{\romannumeral 1} of the vocal data based on alignment a. Model C is very similar, except based on alignment b which is obtained with Model B, etc. Another four sets of Models E, F, G, H are below. For example Model E includes models E-1, 2, 3, 4, Models F,G and H include F-1,2 , G-1,2,3, and H-1,2,3.

We take Model E-4 with fragment-level adaptation within model E as the example. Here every fragment of song (10-35 sec long) was used to train a distinct fragment-level fMLLR matrix, with which Feature \uppercase\expandafter{\romannumeral 3} was obtained. Using all these fragment-level fMLLR features, a single Model E-4 was trained with Alignment d. Similarly for Models E-1, 2, 3 on genre, singer and song levels. The fragment-level Model E-4 turned out to be the best in model E in the experiments.

\subsection{DNN, BLSTM and TDNN-LSTM}
\label{subsec:deep learning models}
 The deep learning models (Models F,G,H) are based on alignment e, produced by the best GMM-HMM model. Models F-1,2 are respectively for regular DNN and multi-target, LibriSpeech phonemes and vocal data phonemes taken as two targets. The latter tried to adapt the speech model to the vocal model, with the first several layers shared, while the final layers separated.
 
 Data augmentation with speed perturbation\cite{ko2015audio} was implemented in Models G, H to increase the quantity of training data and deal with the problem of changing singing rates. For 3-fold, two copies of extra training data were obtained by modifying the audio speed by 0.9 and 1.1. For 5-fold, the speed factors were empirically obtained as 0.9, 0.95, 1.05, 1.1.  1-fold means the original training data. 

 Models G-1,2,3 used projected LSTM (LSTMP)\cite{sak2014long} with 40 dimension MFCCs and 50 dimension i-vectors with output delay of 50ms. BLSTMs were used at 1-fold, 3-fold and 5-fold. 

 Models H-1,2,3 used TDNN-LSTM\cite{peddinti2017low}, also at 1-fold, 3-fold and 5-fold, with the same features as Model G.

\subsection{Special Approaches for Prolonged Vowels}
\label{subsec:lex}
Consider the many errors caused by  the frequently appearing prolonged vowels in song audio, we considered two approaches below.

\subsubsection{Extended Lexicon}
\label{subsubsec:exten_lex}

The previously proposed approach \cite{kawai2016speech} was adopted here as shown by the example in Fig.\ref{fig:extened-vowels}(a). For the word ``apple'', each vowel within the word ( but not the consonants) can be either repeated or not, so for a word with $n$ vowels, $2^{n}$ pronunciations become possible. In the experiments below, we only did it for words with $n\leq3$.

\subsubsection{Increased Self-looped Transition Probabilities}
\label{subsubsec:self_loop}

\begin{figure}[]
\begin{minipage}[b]{1.0\linewidth}
  \centering
  \centerline{\includegraphics[width=8.5cm]{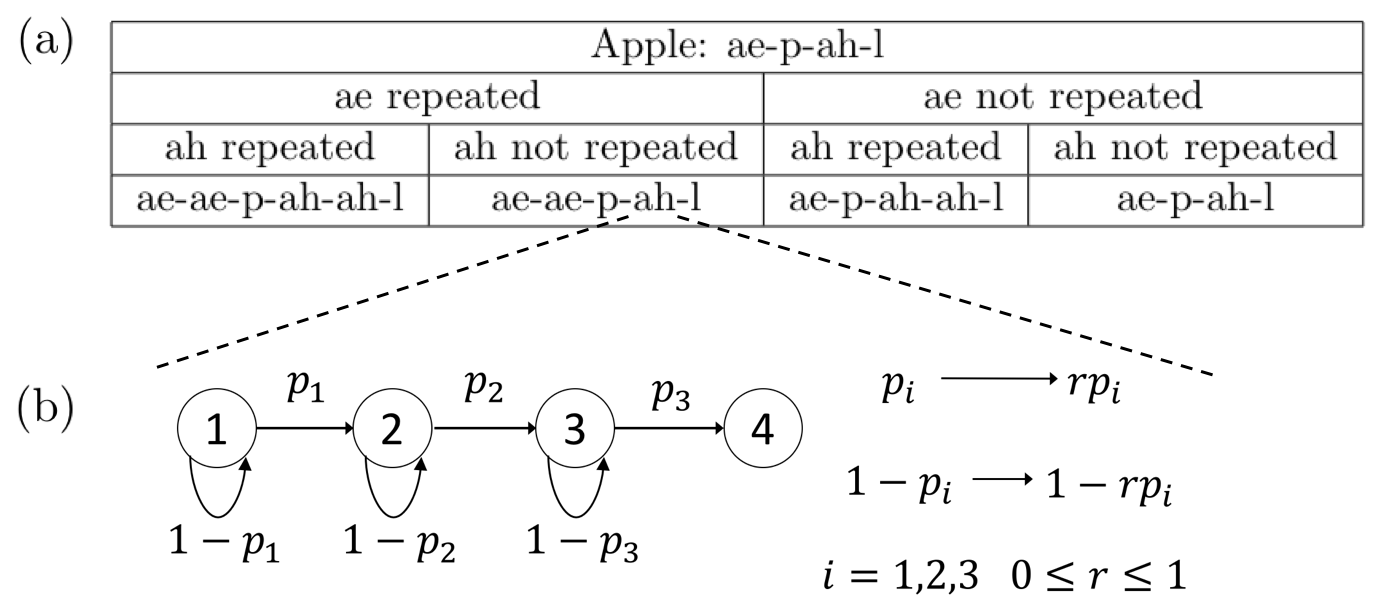}}
\end{minipage}
\caption{Approaches for prolonged vowels: (a) extended lexicon (vowels can be repeated or not), (b) increased self-loop transition probabilities (transition probabilities to the next state reduced by $r$).}

\label{fig:extened-vowels}
\end{figure}

This is also shown in Fig.\ref{fig:extened-vowels}. Assume an vowel HMM have $m+1$ states (including an end state). Let the original self-looped probability of state $i$ is denoted $1-p_{i}$ and the probability of transition to the next state is $p_{i},\ i=1,2,...,m$. We increased the self-looped transition probabilities by replacing $p_{i}$ by $rp_{i}$. This was also done for vowel HMMs only but not for consonants.

\section{EXPERIMENTS}
\label{sec:exp}

\subsection{Data Analysis}
\begin{figure}[htb]

\begin{minipage}[b]{1.0\linewidth}
  \centering
  \centerline{\includegraphics[width=7.5cm]{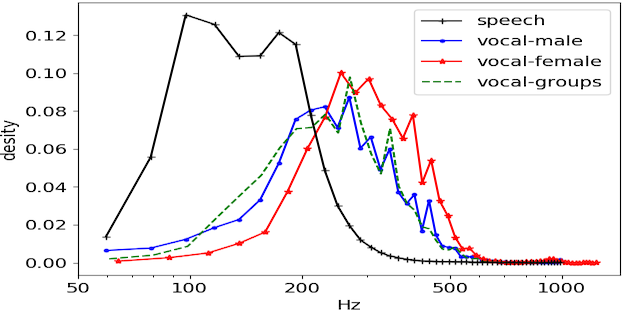}}
\end{minipage}
\caption{Histogram of pitch distribution.}
\label{fig:pitch dist.}
\end{figure}
\subsubsection{Language Model (LM) statistics}

We analyzed the perplexity and out-of-vocabulary(OOV) rate of the two language models (trained with LibriSpeech and Lyrics respectively) tested on the transcriptions of the testing set of vocal data. Both models are  3-gram, pruned with SRILM with the same threshold. LM trained with lyrics was found to have a significantly lower perplexity(123.92 vs 502.06) and a much lower OOV rate (0.55\% vs 1.56\%).

\begin{table}[htb]
\centering
\begin{tabular}{|c|c|l|c|c|}
\hline
\multicolumn{2}{|c|}{} & Acoustic Models & WER(\%) & PER(\%)\\
\hline
\multicolumn{2}{|c|}{\multirow{2}{*}{  \rotatebox{90}{
\begin{tabular}[c]{@{}c@{}}
Libri\\Speech \\ LM
\end{tabular} } }} &
\begin{tabular}[c]{@{}l@{}}
(1) Model A:\\LibriSpeech(SAT)
\end{tabular} & 96.21 & 87.17\\
\cline{3-5}
\multicolumn{2}{|c|}{} & \begin{tabular}[c]{@{}l@{}}
(2) Model E-4:\\fragment-level
\end{tabular} & 88.26 & 77.18\\
\hline
\multicolumn{2}{|c|}{} & \begin{tabular}[c]{@{}l@{}}
(3) Model E-4:\\fragment-level
\end{tabular} & 80.40 & 68.80\\
\cline{2-5}
\multirow{9}{*}{  \rotatebox{90}{
\begin{tabular}[c]{@{}c@{}}
Lyrics Language Model
\end{tabular} } } &
\multirow{9}{*}{  \rotatebox{90}{
\begin{tabular}[c]{@{}c@{}}
Extended Lexicon
\end{tabular} } } &
\begin{tabular}[c]{@{}l@{}}
(4) Model B:\\Monophone
\end{tabular} & 86.57 & 76.10\\
\cline{3-5}
& & \begin{tabular}[c]{@{}l@{}}
(5) Model C:\\Triphone
\end{tabular} & 81.58 & 71.11\\
\cline{3-5}
& & \begin{tabular}[c]{@{}l@{}}
(6) Model D:\\Triphone
\end{tabular} & 82.02 & 72.10\\
\cline{3-5}
& & \begin{tabular}[c]{@{}l@{}}
(7) Model E-4:\\fragment-level
\end{tabular} & 77.08 & 66.04\\
\cline{3-5}
& & \begin{tabular}[c]{@{}l@{}}
(8) Model E-4:\\fragment-level\\
+Increased Trans. Prob.
\end{tabular} & 76.62 & 65.79\\
\cline{3-5}
& & \begin{tabular}[c]{@{}l@{}}
(9) Model F-1\\DNN (regular)
\end{tabular} & 75.56 & 65.64\\
\cline{3-5}
& & \begin{tabular}[c]{@{}l@{}}
(10) Model F-2\\DNN (multi-target)
\end{tabular} & 75.84 & 65.56\\
\cline{3-5}
& & \begin{tabular}[c]{@{}l@{}}
(11) Model G-1\\BLSTM (1-fold)
\end{tabular} & 79.94 & 70.27\\
\cline{3-5}
& & \begin{tabular}[c]{@{}l@{}}
(12) Model G-2\\BLSTM (3-fold)
\end{tabular} & 74.32 &  63.86 \\
\cline{3-5}
& & \begin{tabular}[c]{@{}l@{}}
(13) Model G-3\\BLSTM (5-fold)
\end{tabular} & 75.35 & 65.50\\
\cline{3-5}
& & \begin{tabular}[c]{@{}l@{}}
(14) Model H-1\\TDNN-LSTM (1-fold)\\
\end{tabular} & 79.01 & 69.20\\
\cline{3-5}
& & \begin{tabular}[c]{@{}l@{}}
(15) Model H-2\\TDNN-LSTM (3-fold)\\
\end{tabular} & {\bf 73.90} & 64.33\\
\cline{3-5}
& & \begin{tabular}[c]{@{}l@{}}
(16) Model H-3\\TDNN-LSTM (5-fold)\\
\end{tabular} & 74.53 & {\bf 63.70 }\\
\hline
\end{tabular}
\caption{Word error rate (WER) and phone error rate (PER) over the test set of vocal data.}
\label{tab: all results}
\end{table}

\begin{figure*}[tb]
  \centering
  \centerline{\includegraphics[width=\textwidth]{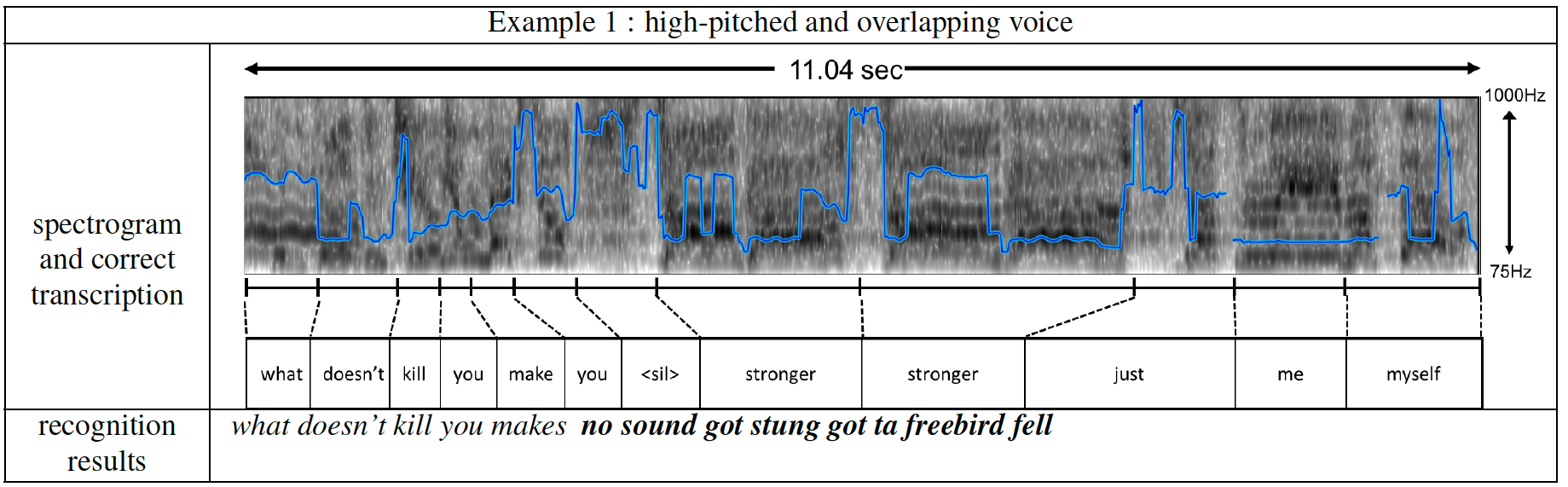}}
\caption{Sample recognition errors produced by Model E-4 : fragment-level in row(7) of Table.\ref{tab: all results}.}
\label{fig:error analysis}
\end{figure*}

\subsubsection{Pitch Distribution}
Fig.\ref{fig:pitch dist.} depicts the histogram for pitch distribution for speech and different genders of vocal. We can see the pitch values of vocal are significantly higher with a much wider range, and female singers produce slightly higher pitch values than male singers and groups.

\subsection{Recognition Results}
The primary recognition results are listed in Table.\ref{tab: all results}. Word error rate (WER) is taken as the major performance measure, while phone error rate (PER) is also listed as references. Rows (1)(2) on the top are for the language model trained with LibriSpeech data, while rows (3)-(16) for the language model trained with lyrics corpus. In addition, in rows (4)-(16) the lexicon was extended with possible repetition of vowels as explained in subsection \ref{subsubsec:exten_lex}. Rows (1)-(8) are for GMM-HMM only, while rows (9)-(16) with DNNs, BLSTMs and TDNN-LSTMs.

Row(1) is for Model A in Fig.\ref{fig:approach flow} taken as the baseline, which was trained on LibriSpeech data with SAT, together with the language model also trained with LibriSpeech. The extremely high WER (96.21\%) indicated the wide mismatch between speech and song audio, and the high difficulties in transcribing song audio. This is taken as the baseline of this work. After going through the series of Alignments a, b, c, d and training the series of Models B, C, D, we finally obtained the best GMM-HMM model, Model E-4 in Model E with fMLLR on the fragment level, as explained in section \ref{sec:approach} and shown in Fig.\ref{fig:approach flow}. As shown in row(2) of Table.\ref{tab: all results}, with the same LibriSpeech LM, Model E-4 reduced WER to 88.26\%, and brought an absolute improvement of 7.95\% (rows (2) vs. (1)), which shows the achievements by the series of GMM-HMM alone. When we replaced the LibriSpeech language model with Lyrics language model but with the same Model E-4, we obtained an WER of 80.40\% or an absolute improvement of 7.86\% (rows (3) vs. (2)). This shows the achievement by the Lyrics language model alone.

We then substituted the normal lexicon with the extended one (with vowels repeated or not as described in subsection \ref{subsubsec:exten_lex}), while using exactly the same model E-4, the WER of 77.08\% in row (7) indicated the extended lexicon alone brought an absolute improvement of 3.32\% (rows (7) vs. (3)). Furthermore, the increased self-looped transition probability ($r=0.9$) in subsection \ref{subsubsec:self_loop} for vowel HMMs also brought an 0.46\% improvement when applied on top of the extended lexicon (rows (8) vs. (7)). The results show that prolonged vowels did cause problems in recognition, and the proposed approaches did help.

Rows (4)(5)(6) for Models B, C, D show the incremental improvements when training the acoustic models with a series of improved alignments a, b, c, which led to the Model E-4 in row (7). Some preliminary tests with p-norm DNN with varying parameters were then performed. The best results for the moment were obtained with 4 hidden layers, 600 and 150 hidden units for p-norm nonlinearity\cite{zhang2014improving}. The result in rows (9) shows absolute improvements of 1.52\% (row (9) for Model F-1 vs. row (7)) for regular DNN. Rows(10) is for Models F-1 DNN (multi-target).  

Rows (11)(12)(13) show the results of BLSTMs with different factors of data augmentation described in \ref{subsec:deep learning models}. Models G-1,2,3 used three layers with 400 hidden states and 100 units for recurrent and projection layer, however, since the amount of training data were different, the number of training epoches were 15, 7 and 5 respectively. Data augmentation brought much improvement of 5.62\% (rows (12) v.s.(11)), while 3-fold BLSTM outperformed 5-fold by 1.03\%. Trend for Model H (rows (14)(15)(16)) is the same as Model G, 3-fold turned out to be the best. Row (15) of Model TDNN-LSTM achieved the lowest WER(\%) of 73.90\%, with architecture $T^{130}T^{130}L^{130}T^{520}T^{520}L^{130}T^{520}T^{520}L^{130}$, while $T^{n}$ and $L^{m}$ denotes that the size of TDNN layer was $n$ and the size of hidden units of forward LSTM was $m$. The WER achieved here are relatively high, indicating the difficulties and the need for further research.

\subsection{Different Levels of fMLLR Adaptation}

In Fig.\ref{fig:approach flow} Model E includes different models obtained with fMLLR over different levels, Models E-1,2,3,4. But in Table.\ref{tab: all results} only Model E-4 is listed. Complete results for Models E-1,2,3,4 are listed in Table.\ref{tab: fMLLR GMM-HMM}, all for Lyrics Language Model with extended lexicon. Row (4) here is for Model E-4, or fMLLR over fragment level,  exactly row (7) of Table.\ref{tab: all results}. Rows (1)(2)(3) are the same as row (5) here, except over levels of genre, singer and song. We see fragment level is the best, probably because fragment(10-35 sec long) is the smallest unit and the acoustic characteristic of signals within a fragment is almost uniform (same genre, same singer and the same song).

\begin{table}[]
\centering
\begin{tabular}{|c|l|c|c|}
\hline
 & Acoustic Model & WER(\%) & PER(\%) \\
\hline
\multirow{5}{*}{  \rotatebox{90}{
\begin{tabular}[c]{@{}l@{}}
Lyrics \\Language Model\\
Extended Lexicon
\end{tabular} } } 
& \begin{tabular}[c]{@{}c@{}}
 (1) Model E-1,\\
 genre-level 
\end{tabular} & 84.24 & 68.92 \\
\cline{2-4}
& \begin{tabular}[c]{@{}c@{}}
(2) Model E-2, \\
singer-level 
\end{tabular} & 78.53 & 68.48\\
\cline{2-4}
& \begin{tabular}[c]{@{}c@{}}
(3) Model E-3, \\
song-level 
\end{tabular} & 78.80 & 68.24\\
\cline{2-4}
& \begin{tabular}[c]{@{}c@{}}
(4) Model E-4, \\
fragment-level 
\end{tabular} & \textbf{77.08} & \textbf{66.04}\\
\hline
\end{tabular}
\caption{Model E : GMM-HMM with fMLLR over different levels.}
\label{tab: fMLLR GMM-HMM}
\end{table}

\subsection{Error Analysis}

From the data, we found errors frequently occurred under some specific circumstances, such as high-pitched voice, widely varying phone duration, overlapping verses (multiple people sing simultaneously), and residual background music.

Figure \ref{fig:error analysis} shows a sample recognition results obtained with Model E-4 as in row(7) of Table.\ref{tab: all results}, showing the error caused by high-pitched voice and overlapping verses. At first, the model successfully decoded the words, \textit{"what doesn't kill you makes"}, but afterward the pitch went high and a lower pitch harmony was added, the recognition results then went totally wrong.

\section{Conclusion}
\label{sec:conclusion}
In this paper we report some initial results of transcribing lyrics from commercial song audio using different sets of acoustic models, adaptation approaches, language models and lexicons. Techniques for special characteristics of song audio were considered. The achieved WER was relatively high compared to experiences in speech recognition. However, considering the much more difficult problems in song audio and the wide difference between speech and singing voice, the results here may serve as good references for future work to be continued.


\bibliographystyle{IEEEbib}

\nocite{*}

\end{document}